\documentclass[aps,jcp,preprint,superscriptaddress,letterpaper]{revtex4}

\usepackage[utf8]{inputenc}

\usepackage{amsmath, amssymb}
\usepackage{amsfonts}
\usepackage{tabularx,booktabs,array,dcolumn}
\usepackage{setspace}
\usepackage{vmargin}
\usepackage{graphicx,psfrag,subfigure}
\usepackage{xcolor}

\usepackage{marvosym}

\usepackage[all]{xy}

\usepackage{tikz}
\usetikzlibrary{calc}
\usetikzlibrary{patterns}

\usepackage{threeparttable}
\usepackage{rotating}
\usepackage{multirow}

\newcommand{\mx}[1]{\boldsymbol{#1}}

\newcommand{\bm}[1]{\boldsymbol{#1}}

\newcommand{\cm}{cm$^{-1}$}

\def\Eh{$\text{E}_\text{h}$}

\def\tr{^{\text{T}}}

\def\iim{\text{i}}

\def\npart{N_\text{p}}
\def\sp{S_\text{p}}
\def\se{S_\text{e}}

\begin{document}

\title{
H$_3^+$ as a five-body problem described with 
explicitly correlated Gaussian basis sets
}
\date{\today}

\author{Andrea Muolo}
\affiliation{ETH Z\"urich, Laboratory of Physical Chemistry, Vladimir-Prelog-Weg 2, 8093 Z\"urich, Switzerland}

\author{Edit M\'atyus}
\email{matyuse@caesar.elte.hu}
\affiliation{Institute of Chemistry, ELTE, E\"otv\"os Lor\'and University, P\'azm\'any P\'eter s\'et\'any 1/A, 1117 Budapest, Hungary}

\author{Markus Reiher}
\email{markus.reiher@phys.chem.ethz.ch}
\affiliation{ETH Z\"urich, Laboratory of Physical Chemistry, Vladimir-Prelog-Weg 2, 8093 Z\"urich, Switzerland}

\begin{abstract}
\noindent %
Various explicitly correlated Gaussian (ECG) basis sets are considered for the solution
of the molecular Schrödinger equation with particular attention to the simplest polyatomic system,
H$_3^+$.
Shortcomings and advantages are discussed for plain ECGs, ECGs with the global vector representation, 
floating ECGs and their numerical projection, and ECGs with complex parameters.
The discussion is accompanied with particle density plots to visualize the observations.
In order to be able to use large complex ECG basis sets in molecular calculations, a numerically stable 
algorithm is developed, the efficiency of which is demonstrated 
for the lowest rotationally and vibrationally excited states of H$_2$ and H$_3^+$. 
\end{abstract}

\maketitle

\section{Introduction}
\noindent%
Recent progress in the experimental energy resolution \cite{ChHuNi18,MaMc19} of spectroscopic transitions
of small molecules urges theoretical and computational methods to deliver orders of magnitude more accurate 
molecular energies than ever before. The current and near future energy resolution 
of experiments allow for a direct assessment of relativistic quantum electrodynamics effects 
and beyond them,
as soon as calculations with a low uncertainty
become available. For small molecules, composed from just a few electrons and a few nuclei, this endavour 
should be realistic within the near future. 
A remarkable experiment--theory concourse 
has been unfolding for the 
three-particle H$_2^+$ molecular ion \cite{KoHiKa17,AlHaKoSc18} 
and for the four-particle H$_2$ molecule \cite{ChHuNi18,WaYa18,PuKoCzPa19,HoBeSaEiUbJuMe19}. 
In addition, there are
promising initial results for the five-particle He$_2^+$ \cite{TuPaAd12,SeJaMe16,Ma18he2p,JaSeMe18}
for which an explicit five-particle treatment, 
at least for the lowest vibrational and rotational excitations,  
should be possible \cite{StBuAd09}.

H$_3^+$ is also a five-particle system, but it is a polyatomic system.  
In comparison with atoms and diatomic molecules, there has been very little 
progress achieved for polyatomics over the past two decades 
regarding an accurate description of the coupled quantum mechanical motion of the electrons
and the atomic nuclei.
In addition to the variational treatment considered in the present work, 
non-adiabatic perturbation theory offers an alternative route for
closing the gap between theory and experiment. The single-state non-adiabatic Hamiltonian has been 
know for a long time \cite{HeAs66,HeOg98,BuMo77,BuMo80,PaKo09,prx17} 
and has been used a few times in practice \cite{Sch01H2O,PrCeJeSz17,Ma18he2p},
while the general working equations for the effective non-adiabatic nuclear Hamiltonian
for multiple, coupled electronic states have been formulated only recently \cite{MaTe19}.

We have already worked on the development of explicitly correlated Gaussian (ECG) ans\"atze 
in relation with the variational solution of polyatomics (electrons plus nuclei). Last year, 
we proposed to use (numerically) projected floating ECGs, which allowed us to approach 
the best estimate obtained on a potential energy surface (PES) for the Pauli-allowed ground state
within 70~\cm\ (31~\cm\ with basis set extrapolation) \cite{MuMaRe18proj}.

The present work starts with an overview of the advantages and shortcomings of the different 
ECG representations together with proton density plots which highlight important qualitative features.
Then, we develop an algorithm which ensures a numerically stable
variational optimization of extensive sets of ECGs with complex parameters, another promising ansatz
for molecular calculations \cite{BuAd06,BuAd08}, and demonstrate its applicability for 
the lowest rotational and vibrational states of H$_2$ and H$_3^+$.

\section{Explicitly correlated Gaussians}
\noindent 
We consider the solution of the time-independent Schrödinger equation (in Hartree atomic units)
including all electrons and atomic nuclei, in total $\npart$ particles, of the molecule, 
\begin{align}
  \left[%
    -\sum_{i=1}^{\npart} \frac{1}{2m_i}\Delta_{\mx{r}_i}
    +\sum_{i=1}^{\npart}\sum_{j>i}^{\npart}
      \frac{Z_iZ_j}{|\mx{r}_i-\mx{r}_j|}
  \right]
  \Psi 
  =
  E\Psi \; .
\end{align}
with electric charges $Z_i$, $Z_j$ and positions $\mx{r}_i$, $\mx{r}_j$.
The exact quantum numbers of the molecular energies and wave functions, $E$ and $\Psi$, 
are the total angular momentum quantum numbers, $N$ and $M_N$, 
the parity, $p$,
and the spin quantum numbers for each particle type, $S_a, M_{S_a},S_b, M_{S_b},\ldots$.

We obtain increasingly accurate approximations to 
the $\Psi$ molecular wave function by using a linear combination
of anti-symmetrized products of (many-particle) spatial, 
$\psi_i^{[N,M_N,p,]}$, 
and spin, $\chi_i^{[S_a,M_{S_a},S_b,M_{S_b},\ldots]}$, functions, 
\begin{align}
  \Psi^{[N,M_N,p, S_a,M_{S_a},S_b,M_{S_b},\ldots]} 
  = 
  \sum_{i=1}^{N_\text{b}} 
  c_i 
  \hat{\mathcal{A}}\lbrace \psi_i^{[N,M_N,p]}  \chi_i^{[S_a,M_{S_a},S_b,M_{S_b},\ldots]} \rbrace \; ,
\end{align}
where $N_\text{b}$ is the number of basis functions and 
$\hat{\mathcal{A}}$ is the anti-symmetrization operator for fermions 
(we would need to symmetrize the product for bosonic particles).
The non-linear parameters of the spatial and the spin functions 
are optimized based on the variational principle \cite{SuVaBook98,rmp13} and
the $c_i$ coefficients are determined by solving the linear variational
problem in a given basis set.

Concerning the construction of the basis set, 
explicitly correlated Gaussian (ECG) functions have been successfully used as spatial basis functions 
for a variety of chemical and physical problems \cite{chemrev13,rmp13}.
In what follows, we consider various ECG basis sets aiming at an accurate solution 
approaching spectroscopic accuracy \cite{specacc} of the molecular Schrödinger equation.
A precise description of vibrational states of di- and polyatomic molecules assumes the use of basis functions 
which have sufficient flexibility to describe the nodes of the wave function along the interparticle distances, 
sharp peaks corresponding to the localization of the nuclei displaced from the center of mass,
and allow us to obtain efficiently the solutions corresponding to the exact quantum numbers of 
this non-relativistic problem.

Concerning the spin functions,
we use the spin functions of two and three identical spin-1/2 fermions (electrons and protons)
with the spin quantum numbers $(S,M_S)=(0,0)$ and $(S,M_S)=(1/2,1/2)$, respectively,  
formulated according to Refs.~\cite{MaRe12,SuVaBook98}. 

In the case of H$_3^+$, the mathematically lowest-energy
(ground electronic, zero-point vibrational) state of the Schrödinger equation
with $N=0$ and $p=+1$ is not allowed by the Pauli principle (for the $S_\text{e}=0$
electrons' and $S_\text{p}=1/2$ protons' spin states), or 
in short, the non-rotating vibrational and electronic ground state of H$_3^+$ 
is spin forbidden \cite{LiMc01,BuJe98}.
The lowest-energy, Pauli-allowed state
is the vibrational ground state ($v=0$) with $N=1$ and $p=-1$ 
(the first rotationally excited state). 
The lowest-energy state with $N=0$ is 
the $(0,1^1)$ fundamental vibration \cite{LiMc01},
which corresponds to asymmetric distortions (anti symmetric for the proton exchange) 
with respect to the equilateral triangular equilibrium structure.

For the assessment and visualization
of the results obtained with the different spatial basis sets,
we use particle density functions, which are very
useful in analyzing the qualitative properties 
of the molecular wave function \cite{MaHuMuRe11a,MaHuMuRe11b,Sc19}. 
We will focus on properties of the proton (p) density (measured from the center-of-mass, CM, position): 
\begin{align}
  D_{p,\text{CM}}(\mx{R})
  =
  \langle
    \Psi |
    \delta(\mx{r}_p - \mx{r}_{\text{CM}} -\mx{R})
    |\Psi
  \rangle \; .
\end{align}

\subsection{Plain ECG, polynomial ECG, and ECG-GVR} 
\begin{figure}
  \includegraphics[scale=1.]{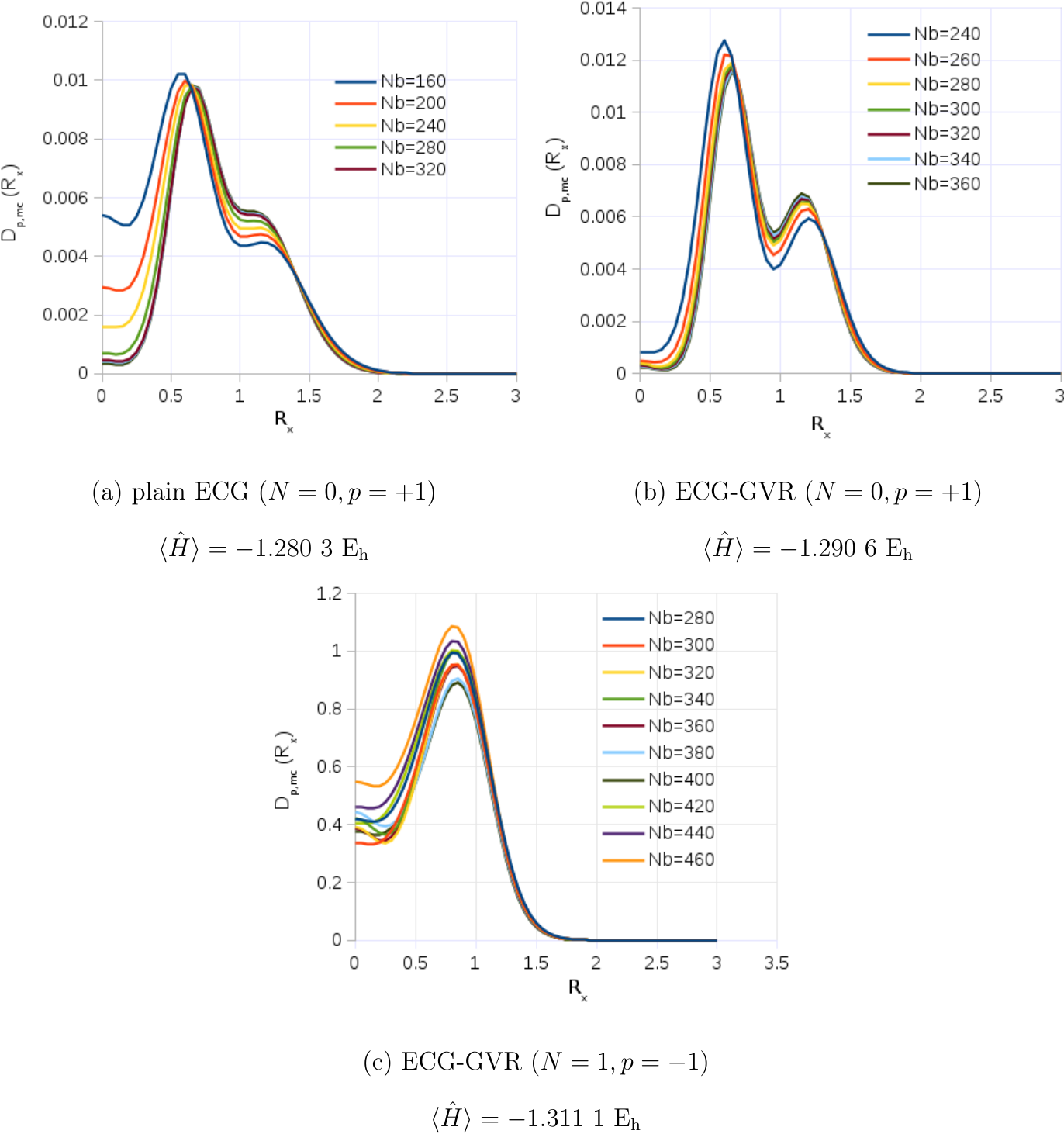} \\
  \caption{%
  Proton probability density 
  for H$_3^+=\lbrace \text{p}^+ ,\text{p}^+,\text{p}^+,\text{e}^-,\text{e}^- \rbrace$ 
  ($S_\text{e}=0$, $S_\text{p}=1/2)$ 
  obtained with plain ECG and ECG-GVR functions 
  (the density is shown along a ray measured from the center of mass). 
  In the ECG-GVR calculations the maximal order for the polynomial prefactor was $2K_\text{max}=20$.
  The reference energy values obtained by us on a PES 
  are $E_\text{PES}(N=0,p=+1)=-1.311\ 950$~\Eh\ and $E_\text{PES}(N=1,p=-1)=-1.323\ 146$~\Eh\
  \cite{pesvalues} (note that the latter value corresponds to the lowest-energy Pauli-allowed state). 
  All quantities in the figure are given in atomic units.
  \label{fig:ecggvr}
  }
\end{figure}

Plain ECG-type functions, 
\begin{align}
  \psi_{\text{ECG}}(\mx{A};\mx{r})
  =
  \exp\left[
    -\mx{r}^\text{T} (\mx{A}\otimes I_3) \mx{r}
  \right] \; 
  \label{eq:ecg}
\end{align}
with the $\mx{A}\in\mathbb{R}^{\npart \times \npart}$ symmetric matrix,
have been successfully used to describe atoms and positron-electron complexes 
(with $N=0$ total angular momentum quantum number and $p=+1$ parity) \cite{SuVaBook98}.
To describe the localization and vibrational excitation of atomic nuclei, 
a linear combination of several plain ECG functions is necessary, which makes their use 
in molecular calculations very inefficient. 
The slow convergence of plain ECGs for the lowest-energy $N=0$ state of H$_3^+$ 
is shown with respect to ECG-GVR (vide infra) in Figure~\ref{fig:ecggvr} (compare sub-Figures~\ref{fig:ecggvr}a 
and \ref{fig:ecggvr}b).

Explicitly correlated Gaussians with the global vector representation (ECG-GVR)
have been originally proposed by Suzuki, Usukura, and Varga in 1998 \cite{SuUsVa98}.
These functions represent a general form of ECGs with polynomial prefactors.
When several ECG-GVR functions are used in a variational procedure, molecular states 
can be converged with a total angular momentum quantum number, $N$, and natural parity, $p=(-1)^N$:
  \begin{align}
    \psi^{[N,M_N]}_{\text{ECG-GVR}}(\mx{r};\mx{A},\mx{{u}},K)
    &=
    Y_{NM_N}(\hat{\mx{v}})\
    |\mx{v}|^{2K+N}\    
    \exp\left[%
     -\mx{r}^\text{T} (\mx{A}\otimes I_3) \mx{r}
    \right] \; ,
      \label{eq:ecggvr}
  \end{align}
  where
  the `global vector' $\mx{v}$ is a linear combination of particle coordinates, 
  \begin{align}
    \mx{v}
    =
    {u}_1 \mx{r}_1 +
    {u}_2 \mx{r}_2 + \ldots + 
    {u}_{\npart} \mx{r}_{\npart}
    =
    (\mx{{u}}\otimes\mx{I}_3)\tr \mx{r} \; ,
  \end{align}
  and $\hat{\mx{v}}$ contains the spherical polar angles corresponding to
  the unit vector $\mx{v}/|\mx{v}|$.

The general ECG-GVR basis set can be very well used to converge the ground- and 
excited states of atoms, positron-electron complexes, as well as 
diatomic molecules (for which plain ECGs would be inefficient) 
with various total angular momentum quantum numbers $N$ 
\cite{ArRiVa05,VaUsSu98,MaRe12,Ma13,FeMa19a}.
It is important to stress, however, 
that higher vibrational excitations, heavier nuclei, or higher $N$ values require 
the use of higher-order polynomials in front of the ECG, which make the integral evaluation
and the entire calculation computationally more demanding.

For $N=0$, an ECG-GVR with the special parameterization $u_i=1$, $u_j=-1$ and $u_k=0$ $(k\neq i,j)$,
simplifies to an ECG with a single (even-power) polynomial prefactor, 
  \begin{align}
    \psi^{[0,0]}_{\text{ECG-r}}(\mx{r};\mx{A},K)
    &=
    |\mx{r}_i-\mx{r}_j|^{2K}\    
    \exp\left[%
     -\mx{r}^\text{T} (\mx{A}\otimes I_3) \mx{r}
    \right] 
      \nonumber    
    \\
    &=
    r_{ij}^{2K}\    
    \exp\left[%
     -\mx{r}^\text{T} (\mx{A}\otimes I_3) \mx{r}
    \right] \; ,
      \label{eq:ecgpref}
  \end{align}
which has been successfully used 
to describe vibrations of diatomic molecules by Adamowicz and co-workers \cite{KiAd99,BuAd03,CaBuAd03}.

In spite of the success of these type of basis functions for atoms and diatoms, the ECG-GVR ansatz
was found to be inefficient \cite{MuMaRe18proj}
(comparable to the single-polynomial ECG ansatz, Eq.~(\ref{eq:ecgpref}) \cite{BeBuAd05})
to converge the five-particle energy of H$_3^+$ within spectroscopic accuracy \cite{specacc}. 
Even the proton density can be hardly converged (Figure~\ref{fig:ecggvr}),
while the energy has uncertainties (much) larger than 1~m\Eh. 
The proton density for the lowest-energy $N=0$ state has two peaks, 
which may be qualitatively correct, since this state  (if converged) corresponds to the anti-symmetric fundamental vibration, 
which should feature two peaks in the proton density
measured from the center of mass. 
The two peaks appear already in the plain ECG calculation (Figure~\ref{fig:ecggvr}a),
but plain ECG densities have even larger uncertainties. 
Further increase of the basis set (towards convergence) is hindered
by near-linear dependency problems, which is an indication of
insufficient flexibility in the mathematical form of the basis functions.

Figure~\ref{fig:ecggvr}c shows the (convergence of the) proton density obtained with
ECG-GVR functions for the lowest-energy rotational ($N=1$) state which corresponds to 
the lowest-energy Pauli-allowed state of the system. 
Notice the significant amount of density at the origin 
(center of mass) and the large deviations of results obtained with different basis set sizes, 
which must be artifacts due to incomplete convergence 
(compare with Figures~\ref{fig:fECGH3p} and \ref{FIG:rotsym3}).
The `best' (lowest) five-particle energy, we obtained with an ECG-GVR representation 
for the lowest-energy state ($N=1$), 
is 1.8~mE$_\text{h}$ larger than the best
estimate on a potential energy surface (PES) \cite{MuMaRe18proj}.

Hence, the slow convergence of the energy and the density in the ECG-GVR ansatz
is related to the fact that these functions are not flexible
enough to efficiently describe the triangular arrangement (and vibrations) 
of the protons in H$_3^+$ and the spherical symmetry of 
the system at the same time \cite{MuMaRe18proj}.
In principle, it would be possible to define ECG-GVRs
with multiple global vectors which could give
a better account of the rotational and the multi-particle clustering effects
in a polyatomic molecule, but the formalism would be very involved.

\subsection{ECGs with three pre-exponential polynomials}
We note in passing that, in 2005, Bednarz, Bubin, and Adamowicz proposed an ECG ansatz
for H$_3^+$ \cite{BeBuAd05},
\begin{align}
  \psi^{[0,0]}_{\text{ECG-3r}}(\mx{r};\mx{A},k_{12},k_{13},k_{23})
  &=
  r_{12}^{2k_{12}}
  r_{13}^{2k_{13}}
  r_{23}^{2k_{23}}  
  \exp\left[%
    -\mx{r}^\text{T} (\mx{A}\otimes I_3) \mx{r}
  \right] \; ,
  \label{eq:ecg3r}
\end{align}
by including polynomial pre-factors for all three proton-proton distances,
$r_{ij}=|\mx{r}_i-\mx{r}_j|$ $(i,j=1,2,3,j>i)$.
The integral expressions have been formulated, 
but to our best knowledge, they have never been used in practical calculations
due to their very complicated form and numerical instabilities \cite{BuFoAd16}.

\subsection{Floating ECGs with explicit projection} 
\begin{figure}
  \includegraphics[scale=1.]{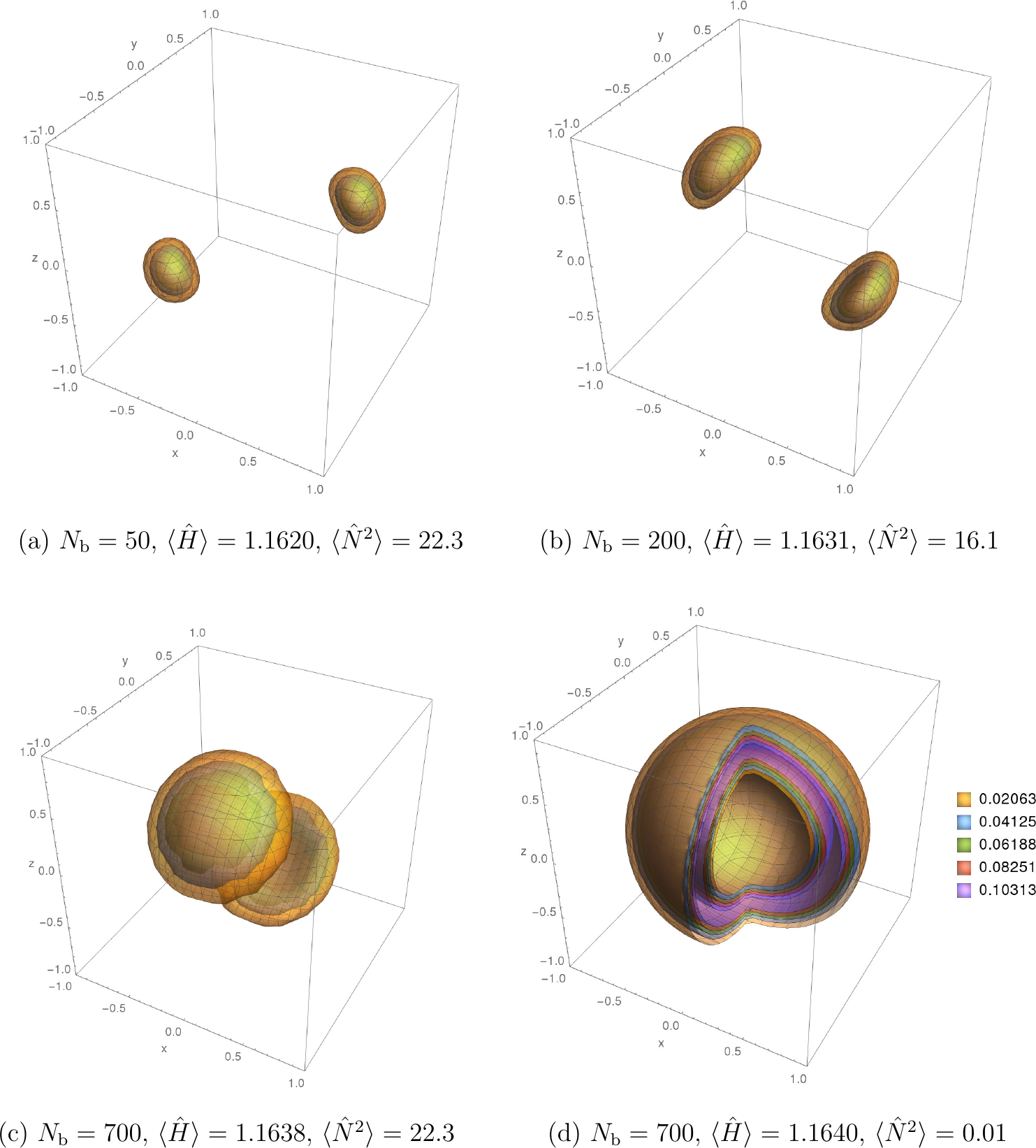}  \\
  \caption{%
    Proton probability density for the ground state of H$_2=\lbrace \text{p}^+,\text{p}^+,\text{e}^-,\text{e}^- \rbrace$
    $(S_\text{e}=0,S_\text{p}=0)$ obtained with floating ECGs. 
    (a)--(c): non-projected (symmetry breaking) FECGs;
    (d)~FECG basis functions numerically projected 
    onto the $(N=0,p=+1)$ irrep of $O(3)$  using 22 quadrature points for each Euler angle \cite{MuMaRe18proj}. 
    $N_\text{b}$ is the number of basis functions.
    The energy and the square of the total angular momentum operator are
    $\langle \hat{H} \rangle=-1.164\,025\,031$~\Eh\ \cite{PaKo09}
    and
    $\langle \hat{N}^2\rangle=0$, respectively.
    All quantities in the figure are given in atomic units.
  \label{fig:fECGH2}}
\end{figure}

\begin{figure}
  \includegraphics[scale=1.]{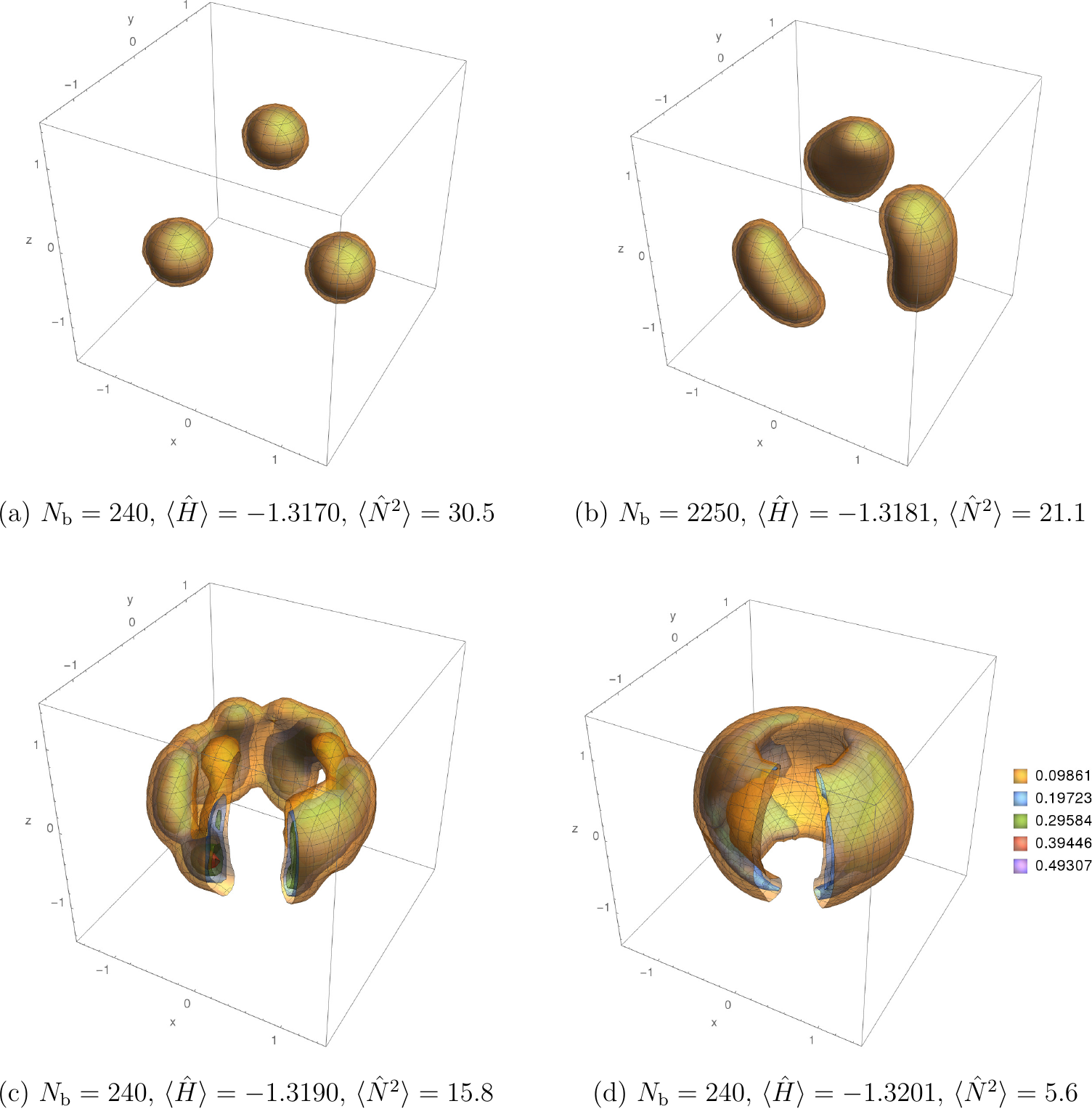}  \\
  \caption{%
    Proton probability density for the lowest-energy, Pauli-allowed
    state of H$_3^+=\lbrace \text{p}^+,\text{p}^+,\text{p}^+,\text{e}^-,\text{e}^-\rbrace$, 
    which is the first rotationally excited state ($N=1$) of the zero-point vibration.
    (a)--(b) non-projected (symmetry breaking) FECGs;
    (c)--(d) FECGs approximately projected onto the $(N=1,p=-1)$ irrep
    of $O(3)$  using 4 and 8 quadrature points for each Euler angle, respectively \cite{MuMaRe18proj}. 
    For the exact wave function $\langle \hat{N}^2\rangle = 2$ ($N=1$).
    All quantities are given in atomic units.
  \label{fig:fECGH3p}}
\end{figure}

Floating ECGs (FECGs), 
\begin{align}
  \psi_{\text{FECG}}(\mx{A},\mx{s};\mx{r})
  =
  \exp\left[
    -(\mx{r}-\mx{s})^\text{T} (\mx{A}\otimes I_3) (\mx{r}-\mx{s})
  \right] \; ,
  \label{eq:fecg}
\end{align}
offer the flexibility to choose (optimize) not only the exponents but also 
the centers, $\mx{s}\in\mathbb{R}^{\npart\times 3}$, 
which allows one to efficiently describe localization
of the heavy particles in polyatomics. 
At the same time, the FECG functions with arbitary, $\mx{s} \neq \mx{0}$, 
centers do not transform as the irreducible representations (irreps)
of the three-dimensional rotation-inversion group, $O(3)$, 
and therefore, they are neither eigenfunctions of the total squared angular momentum
operator, $\hat{N}^2$, nor of space inversion (parity). 
Although these symmetry properties are numerically recovered during the course of 
the variational optimization converging to the exact solution (see 
Figures~\ref{fig:fECGH2}a--c and \ref{fig:fECGH3p}a--b), 
it is extremely inefficient (impractical) for molecular calculations
to recover the continuous symmetry numerically.

In order to speed up the slow convergence in the FECG ansatz due to the broken spatial symmetry, 
we proposed last year \cite{MuMaRe18proj} to project 
the floating ECG basis functions onto irreps of $O(3)$,
\begin{align}
  \psi_{\text{p-FECG}}^{[N,p]}(\mx{A},\mx{s};\mx{r})
  =
  \frac{1}{4\pi} \int &
    \left[D^{(N)}_{M_NM_N}(\Omega)\right]^\ast \hat{R}(\Omega)\  
    \frac{1}{2}(1 + p\cdot \hat{i})
    \nonumber \\
    &\quad\quad 
    \exp\left[
      -(\mx{r}-\mx{s})^\text{T} (\mx{A}\otimes I_3) (\mx{r}-\mx{s})
    \right] \text{d}\Omega\   \; ,
  \label{eq:pFECG} 
\end{align}
where $\Omega$ collects parameterization of the 3-dimensional rotation, 
\emph{e.g.,} in terms of three Euler angles, 
$D^{(N)}_{M_NM_N}(\Omega)$ is the $(M_N,M_N)$th element of the $N$th-order 
Wigner $D$ matrix, and $\hat{R}(\Omega)$ is the corresponding 
three-dimensional rotation
operator. $p$ is the parity, $+1$ or $-1$,
and $\hat{i}$ is the 3-dimensional space-inversion operator.
Both the $\hat{R}$ rotational and the $\hat{i}$ space-inversion operators
act on the particle coordinates, $\mx{r}$, but the mathematical form of 
the ECGs allowed us to translate their action onto the transformation 
of the ECG parameters, $\mx{A}$ and $\mx{s}$ \cite{MaRe12,MuMaRe18proj}.

For the projected basis functions 
the integral expressions of the non-relativistic operators are in general not known analytically,
and for this reason, we have carried out an approximate, numerical projection 
in Ref.~\cite{MuMaRe18proj}.
Using numerically projected floating ECGs we achieved to significantly improve upon
the five-particle energy for H$_3^+$ and to approach the current best estimate 
(on a potential energy surface) for the Pauli-allowed ground state 
within 70~\cm\ (with extrapolation within 31~\cm).

Properties of (unprojected, symmetry-breaking) and (approximately) projected FECGs
are shown for the example of the proton density of H$_2$ and H$_3^+$
in Figures~\ref{fig:fECGH2} and \ref{fig:fECGH3p}, respectively.
At the beginning of an unprojected calculation, 
the proton density first localizes at around three (two) lobes which 
corresponds to the localization of the protons in H$_3^+$ (and H$_2$) 
exhibiting small-amplitude vibrations in a fixed orientation (which can be understood
as a superposition of several eigenstates with different $N,M_N$, and $p$ values). Then, during the course
of the variational increase of the basis set, the spherical symmetry is recovered  
but the triangular (dumbbell-like) relative configuration of the protons in H$_3^+$ (in H$_2$) is also
described within the proton shells (not shown in the figures).
Numerical projection reconstructs the expected spherical symmetry directly, 
without the need of variational optimization, 
as it is shown in Figures~\ref{fig:fECGH3p}c--d and in Figure~\ref{fig:fECGH2}d.

To construct the density plots, we had to evaluate the proton density at several points 
in space, which is demanding for H$_3^+$ with the current projection scheme. For this reason,
the largest basis set used for the density plot (Figure~\ref{fig:fECGH3p}d) 
is smaller than the best one obtained during the convergence of the five-particle 
energy in Ref.~\cite{MuMaRe18proj}.
Projected FECGs are promising candidates for solving H$_3^+$ as a five-particle
problem and we anticipate further progress along this line in the future.

\subsection{Complex ECGs}
In 2006, Bubin and Adamowicz \cite{BuAd06} proposed to use ECGs with complex parameters (CECGs), 
\begin{align}
  \psi_{\text{CECG}}(\mx{C};\mx{r})
  =
  \exp\left[
    -\mx{r}^\text{T} (\mx{C}\otimes I_3) \mx{r}
  \right] \; ,
  \label{eq:cecg}
\end{align}
to describe vibrational ($N=0$, $p=+1$) states of molecules.
$\mx{C}=\mx{A}+\iim\mx{B}\in\mathbb{C}^{\npart \times \npart}$ is a complex-valued matrix 
with the real, symmetric matrices, $\mx{A}$ and $\mx{B}$. To ensure
square integrability, $\psi$ must decay to zero at large distances. Furthermore, 
to have a positive definite $\psi$, $\mx{A}$ must be positive definite.
Most physical operators have very simple integrals in this basis set
and the integrals can be evaluated with a small number of operations 
(\emph{i.e.,} at low computational cost), which does not increase with
increasing the number of nodes of the basis function (unlike for ECG-GVR or polynomial ECG). 
The rich nodal structure of CECGs, introduced by the $\mx{B}$ imaginary part of the exponent,
can be understood through the Euler identity,
$\text{e}^{-(a+\iim b)r^2}=\text{e}^{-ar^2}[\cos(br^2)-\iim\sin(br^2)]$.

In 2008, Bubin and Adamowicz \cite{BuAd08} 
proposed to extend CECGs for computing $N=1$ states of diatomics with
\begin{align}
  \psi_{\text{zCECG}}(\mx{C};\mx{r})
  =
  \rho_z
  \exp\left[
    -\mx{r}^\text{T} (\mx{C}\otimes I_3) \mx{r}
  \right] \; ,
  \label{eq:cecg}
\end{align}
using $\rho_z=(\mx{r}_{n_1}-\mx{r}_{n_2})_z$, which is
the $z$-component of the displacement vector between the two nuclei, $n_1$ and $n_2$. This ansatz yields a very
good description for the first rotationally excited state of a diatomic molecule
(the electrons' contribution to the total angular momentum is almost negligible).

The analytic matrix elements for the overlap, kinetic energy, Coulomb potential energy,
and particle density (together with the energy gradients with respect to 
the $\mx{C}$ matrix parameters) have been derived by Bubin and Adamowicz 
and the expressions can be found in Refs.~\cite{BuAd06,BuAd08}.

Widespread application of the CECG basis-function family is hindered by 
the fact that matrix operations (matrix inversion, etc.)
are more affected by numerical instabilities in finite (double) precision
complex arithmetics  when compared to real arithmetics.

Earlier this year, Varga proposed \cite{Va19} a numerically stable implementation of the CECG 
functions, through real combinations, 
\begin{align}
  \psi_{\text{C-CECG}}(\bm{r};\mx{C}) 
  &=
  \frac{1}{2}
  \big[%
    \psi_{{\text{CECG}}}(\bm{r};\mx{C}) +
    \psi_{{\text{CECG}}}(\bm{r};\mx{C}^\ast) 
  \big] \; , \\
  \psi_{\text{S-CECG}}(\bm{r};\mx{C}) 
  &=
  \frac{1}{2 \iim}
  \big[%
    \psi^{{\text{CECG}}}(\bm{r};\mx{C}) -
    \psi^{{\text{CECG}}}(\bm{r};\mx{C}^\ast) 
  \big] 
\end{align}
with $\mx{C}^\ast$ being the complex conjugate of $\mx{C}$, 
which allowed him to work with real-valued Hamiltonian and overlap matrices. 
Furthermore, he also proposed an imaginary-time propagation approach
to make the optimization of the complex exponent matrices efficient
for the ground state of molecular systems \cite{Va19}.

\section{Algorithm for numerically stable calculations with complex ECGs}
\noindent%
In this section, we present the key elements of 
a numerically stable algorithm that we developed for 
the original (complex) CECG functions. 

Following Eq.~(\ref{eq:cecg}),
we define a new CECG basis function by specifying 
the $\mx{A}$ and 
$\mx{B} \in \mathbb{R}^{\npart \times \npart}$ 
real symmetric matrices, which give the complex symmetric matrix, 
$\bm{C}={\bm{A}}+\iim\,{\bm{B}}$, 
in the exponent of the ECG.
We work in laboratory-fixed Cartesian coordinates (LFCC) \cite{SiMaRe13,MuMaRe18}
and use a multi-channel optimization procedure, \emph{i.e.,} optimize the coefficient matrices 
corresponding to different translationally invariant (TI) Cartesian coordinate coordinate
representations \cite{MuMaRe18}. Owing to the mathematical form of the ECGs, the transformation of the coordinates
can be translated to the transformation of the parameter matrix \cite{MaRe12}.
In all TI representations, the $\mx{A}$ and $\mx{B}$ matrices are block diagonal,
\emph{i.e.,} the TI and the center-of-mass (CM) blocks do not couple. 
To ensure square integrability, we choose
the CM block of $\mx{A}$ to have non-zero values on its diagonal. 
We choose the same non-zero value for all diagonal entries and for each basis function,
the contribution of which is eliminated
(subtracted) during the evaluation of the integrals. With this choice for the real part $\mx{A}$, 
we are free to set the CM block of the imaginary part $\mx{B}$ to zero, 
and $\psi$, of course, remains positive definite (due to the non-vanishing
CM block of $\mx{A}$).

In order to obtain $N=1$ states, we use CECGs multiplied with the $z$
coordinate of a `pseudo-particle'. 
Bubin and Adamowicz used the $z$ component
of the nucleus-nucleus displacement vector in diatomic molecules \cite{BuAd08}.
We do not choose only a single pseudo-particle but pick different particle pairs 
for the different basis functions 
(and possibly several other linear combinations of the particle 
coordinates, inspired by the ECG-GVR idea \cite{VaSuUs98}) 
to ensure that the contribution of each particle pair to the angular momentum is accounted for.
Hence, our general form for complex basis functions, gzCECG, for $N=1,p=-1$ states is
\begin{align}
  \phi_{\text{gzCECG}}(\mx{r};\mx{C},i)
  = 
  \rho^{(i)}_z(\bm{r}) \, \exp\left[ -\bm{r}^T (\mx{C}\otimes \mx{I}_3) \bm{r}\right] ~,
  \label{EQ:zCECG}  
\end{align}
where $\rho^{(i)}_z$ is the $z$ component of the $i$th translationally invariant vector,
formed as a linear combination of the particle coordinates
\begin{align}
  \rho^{(i)}_z(\bm{r})
  = 
  \sum_{j=1}^{\npart}
    u^{(i)}_j r_{j,z}   
 \; .
 \label{eq:rhoi}
\end{align}
Of course, there are infinitely
many such combinations. In the present calculations, we have included
all possible pairs of particles, \emph{i.e.,} $i$ in Eq.~(\ref{eq:rhoi}) cycles through 
the possible particle pairs only.
For example, there are 
$\left( \begin{array}{c} 5 \\ 2 \end{array} \right) = 10$ 
possible particle pairs in H$_3^+$, and we consider the following 
$\rho_z^{(i)}$-parameterization ($i=1,2,\ldots,10$) in the gzCECG representation:
\begin{enumerate}
  \item[(1)]
    $\mx{u}^{(1)}=(1,-1,0,0,0)$ 
  \item[(2)] 
    $\mx{u}^{(2)}=(1,0,-1,0,0)$ 
  \item[($\ldots$)]
  \item[(10)]
    $\mx{u}^{(10)}=(0,0,0,1,-1)$.
\end{enumerate}

A robust and numerically stable implementation of (gz)CECGs has been a challenging task.
The overlap and  Hamiltonian matrix elements are complex and 
the complex generalized eigenvalue problem 
quickly becomes unstable when increasing the size of the basis set 
in a stochastic variational approach.
We have studied the nature of these instabilities 
and have identified two ingredients producing this unstable behavior.

First, an unrestricted optimization of the $\mx{B}$ matrix generates 
increasingly oscillatory functions, and thus the basis function decays slowly 
in the limit $\bm{r}\rightarrow\infty$.
This behavior affects a broad region of the parameter space; 
it happens, whenever the imaginary part $\mx{B}$
dominates the real part $\mx{A}$.

Second, the analytic overlap and Hamiltonian matrix elements require the calculation of
the determinant and the inverse of the complex, symmetric matrix $\mx{C}$,
the evaluation of which suffer from loss of precision in floating-point arithmetics, \emph{i.e.,}
an ill-conditioned matrix is still invertible, but the inversion is numerically unstable.
The quality of the eigenvalues and eigenfunctions of 
the Hamiltonian matrix (with the complex, non-diagonal overlap matrix)
is thereby compromised by ill-conditioned matrices $C$,
an undesired feature which can be identified by repeating the calculations
with higher-precision arithmetics or by monitoring
the range spanned by the eigenvalues of the matrices \footnote{The range of the eigenvalue spectrum is characterized by the so-called condition number.
The $\kappa(\mx{M})$ condition number for an $\mx{M}$ complex, symmetric matrix
is defined  as the ratio of the largest and the smallest eigenvalues of
$(\mx{M}+\mx{M}^\dagger)/2$.}.

Based on these observations, we propose the following conditions
to ensure numerical stability of the variational procedure in finite-precision arithmetics.
During the course of the variational selection and optimization of the basis function
parameters, we monitor
\begin{enumerate}
  \item[(1)]
    the ratio of the diagonal elements of 
    the real and the imaginary parts of $\mx{C}=\mx{A}+\iim\mx{C}$:
    $A_{ii}/B_{ii} < \epsilon_{1i},\ i=1,\ldots,\npart \; ;$
  \item[(2)] 
    the condition number of $\mx{C}$: 
    $\kappa(\mx{C})<\epsilon_2 \; ;$
  \item[(3)]
    the condition number of the (complex symmetric) $\mx{S}$ overlap and 
    the $\mx{H}$ Hamiltonian matrices: 
    $\kappa(\mx{S})<\epsilon_{3S}$ and 
    $\kappa(\mx{H})<\epsilon_{3H} \; .$
\end{enumerate}
For acceptance of a trial basis function as a new basis function
in the basis set these three conditions must be fulfilled in addition to 
minimization of the energy. In this way, the numerical
stability of the computational procedure can be ensured.
For the present calculations, carried out using double precision arithmetics,
we have found that the same $\epsilon_{1i}=\epsilon_2=\epsilon_{3S}=\epsilon_{3H}
=\epsilon = 10^{10}$ value for each condition 
ensures numerical stability for the desired precision, \emph{i.e.,} 
6--9 significant digits in the energy.
The conditions (1)--(3) and the selected value of $\epsilon$
have been constantly tested during the calculations
by solving the linear variational problem within the actual basis set
with increased (quadruple and beyond) precision arithmetics.

The first two conditions ensure that the parameter optimization algorithm
avoids the regions which would result in overly oscillatory basis functions at large distances,
while the third condition controls the level of linear dependency within 
the (non-orthogonal) basis set.

The computational bottleneck of the (gz)CECG calculations is related to the solution of the 
generalized complex eigenvalue problem as it was also noted in Ref.~\cite{BuFoAd16}. 
For this reason, we have implemented and used the FEAST eigensolver algorithm \cite{Po09},
which is a novel, powerful iterative eigensolver for the generalized, complex, symmetric eigenvalue problem.

\begin{figure}[h]
  \centering
  \includegraphics[scale=1.]{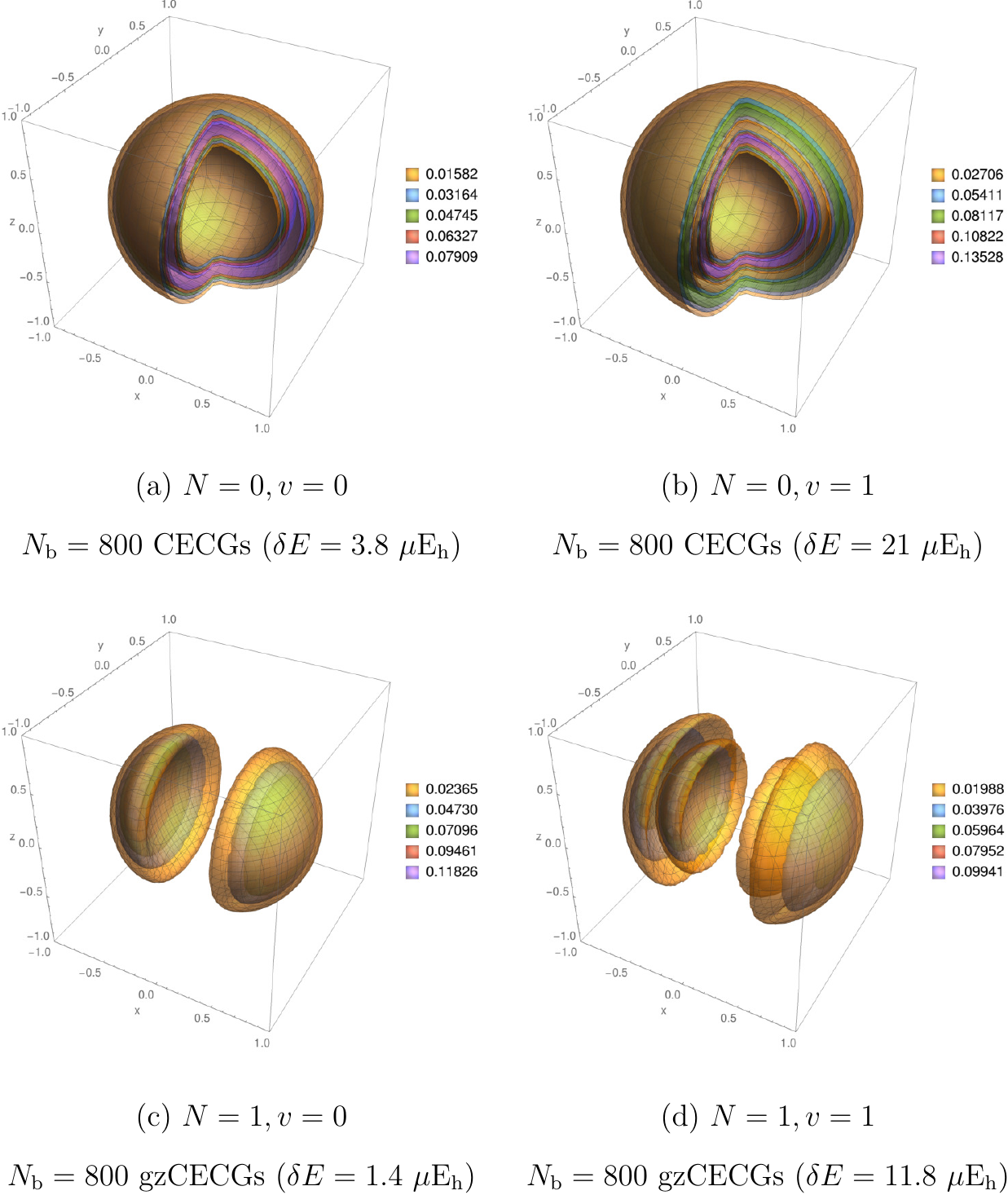}  \\  
  \hfill
  \caption{\label{FIG:rotsym1}
    {%
    Proton probability density calculated for 
    the ground ($v=0,N=0$) and the lowest rotationally ($N=1$) and vibrationally ($v=1$)
    excited states of H$_2=\lbrace \text{p}^+,\text{p}^+,\text{e}^-,\text{e}^-\rbrace$ 
    ($\se=0,\sp=(1-N)/2,p=(-1)^N$) using (gz)CECG functions. 
    The particle densities are converged within figure resolution, deviation
    of the energy from benchmark values \cite{PaKo09} is given in parentheses. 
    All quantities in the figure are given in atomic units.
  }}
\end{figure}

Figure~\ref{FIG:rotsym1} shows the convergence of the proton density (the energy is also given)
for the ground and rotationally and vibrationally excited states of the H$_2$ molecule ($\se=0,\sp=0$).
These results were obtained within a few days on a multi-core workstation. 
While the densities are very well converged, the energies can be further improved by subsequent basis-set 
optimization.

\begin{figure}[h]
  \centering
  \includegraphics[scale=1.]{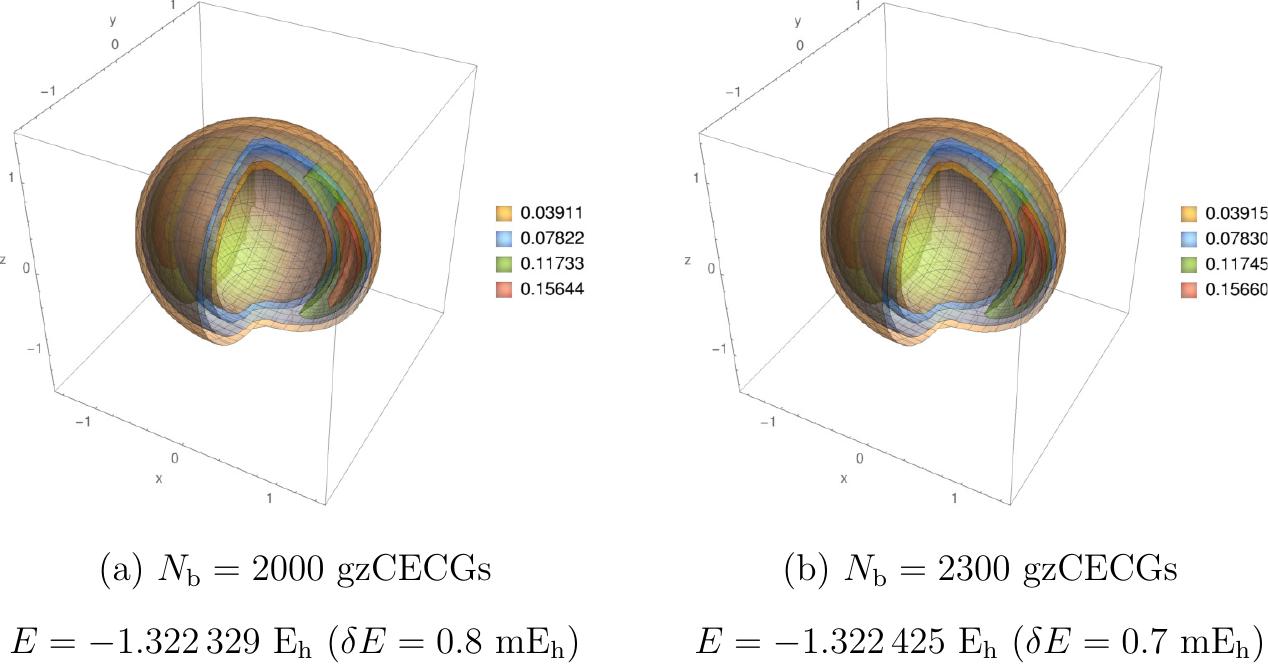}  \\  
  \caption{%
    Proton probability density 
    for the lowest-energy, Pauli-allowed state of H$_3^+$ 
    $(N=1,p=-1,\se=0,\sp=1/2)$
    obtained with gzCECG functions.
    The density is converged within figure resolution (compare plots a and b), while 
    the deviation of the five-particle energy from 
    the best value obtained on a PES in our earlier work \cite{MuMaRe18proj},
    $E_{\text{ref}}=-1.323\,146\ \text{E}_\text{h}$ \cite{pesvalues},
    is given in parentheses ($\delta E=E-E_\text{ref}$).
    All quantities in the figure are given in atomic units.
    \label{FIG:rotsym3}
  }    
\end{figure}

Figure~\ref{FIG:rotsym3} shows our best results obtained for selected states of 
H$_3^+$ ($\se=0,\sp=1/2$) using the numerically stable gzCECG implementation developed in this work.
The proton probability density for the lowest-energy, Pauli-allowed state (zero-point vibration, $N=1$)
is well converged, the difference between Figure~\ref{FIG:rotsym3}a
and Figure~\ref{FIG:rotsym3}b can be hardly seen. 
The best energy is 0.7~mE$_\text{h}\approx 153$~\cm\ higher than the reference value
obtained on a PES \cite{pesvalues}.

\clearpage

\clearpage
\section{Conclusions}
\noindent
Explicitly correlated Gaussian basis sets have been an excellent choice when aiming for ultra-precise energies for 
atoms, electron-positron complexes, and diatomic molecules. However, tight convergence of the energy of H$_3^+$,
the simplest polyatomic system, by including all electrons and protons 
in a variational procedure has not been achieved yet.

In this work, we critically assessed explicitly correlated Gaussian (ECG) basis sets for solving 
the molecular (electrons plus nuclei) Schrödinger equation through the study of the convergence of
the energy and the particle (proton) density. These observations will contribute to developments that
will eventually allow for the convergence of the five-particle energy of H$_3^+$ within spectroscopic accuracy, i.e.,
an uncertainty better than 1~\cm\ ($< 5\times 10^{-6}$~\Eh) for the molecular energy.

In 2018, we developed an algorithm for numerically projected floating ECGs \cite{MuMaRe18proj} to compute 
the lowest-energy state of H$_3^+$ in a variational procedure. 
In the present work, we presented a numerically stable algorithm for another promising basis set for solving H$_3^+$,
complex ECGs, which makes it possible to use large basis set sizes in finite precision arithmetics. 
Although projected floating ECGs provided a somewhat lower energy \cite{MuMaRe18proj} than complex ECGs (present work)
so far, it is currently unclear which type of basis set will finally allow one to reach spectroscopic accuracy
for H$_3^+$ treated as a five-particle system.

Reaching and transgressing this level of uncertainty in a variational computation
will make it possible to directly assess effective non-adiabatic mass models 
and to study relativistic and quantum electrodynamics effects in the high resolutions
spectrum. Such calculations are beyond the scope of the present work and therefore deferred to future studies.

\vspace{2.cm}
\paragraph*{Acknowledgement}
\noindent %
This work was supported by ETH Zurich.
EM acknowledges financial support from a PROMYS Grant (no. IZ11Z0\_166525)  
of the Swiss National Science Foundation and 
ETH~Z\"urich for supporting a stay as visiting professor during 2019.


\begin{thebibliography}{55}
\expandafter\ifx\csname natexlab\endcsname\relax\def\natexlab#1{#1}\fi
\expandafter\ifx\csname bibnamefont\endcsname\relax
  \def\bibnamefont#1{#1}\fi
\expandafter\ifx\csname bibfnamefont\endcsname\relax
  \def\bibfnamefont#1{#1}\fi
\expandafter\ifx\csname citenamefont\endcsname\relax
  \def\citenamefont#1{#1}\fi
\expandafter\ifx\csname url\endcsname\relax
  \def\url#1{\texttt{#1}}\fi
\expandafter\ifx\csname urlprefix\endcsname\relax\def\urlprefix{URL }\fi
\providecommand{\bibinfo}[2]{#2}
\providecommand{\eprint}[2][]{\url{#2}}

\bibitem[{\citenamefont{Cheng et~al.}(2018)\citenamefont{Cheng, Hussels, Niu,
  Bethlem, Eikema, Salumbides, Ubachs, Beyer, Hölsch, Agner
  et~al.}}]{ChHuNi18}
\bibinfo{author}{\bibfnamefont{C.-F.} \bibnamefont{Cheng}},
  \bibinfo{author}{\bibfnamefont{J.}~\bibnamefont{Hussels}},
  \bibinfo{author}{\bibfnamefont{M.}~\bibnamefont{Niu}},
  \bibinfo{author}{\bibfnamefont{H.~L.} \bibnamefont{Bethlem}},
  \bibinfo{author}{\bibfnamefont{K.~S.~E.} \bibnamefont{Eikema}},
  \bibinfo{author}{\bibfnamefont{E.~J.} \bibnamefont{Salumbides}},
  \bibinfo{author}{\bibfnamefont{W.}~\bibnamefont{Ubachs}},
  \bibinfo{author}{\bibfnamefont{M.}~\bibnamefont{Beyer}},
  \bibinfo{author}{\bibfnamefont{N.}~\bibnamefont{Hölsch}},
  \bibinfo{author}{\bibfnamefont{J.~A.} \bibnamefont{Agner}},
  \bibnamefont{et~al.}, \bibinfo{journal}{Phys. Rev. Lett.}
  \textbf{\bibinfo{volume}{121}}, \bibinfo{pages}{013001}
  (\bibinfo{year}{2018}).

\bibitem[{\citenamefont{Markus and McCall}(2019)}]{MaMc19}
\bibinfo{author}{\bibfnamefont{C.~R.} \bibnamefont{Markus}} \bibnamefont{and}
  \bibinfo{author}{\bibfnamefont{B.~J.} \bibnamefont{McCall}},
  \bibinfo{journal}{J. Chem. Phys.} \textbf{\bibinfo{volume}{150}},
  \bibinfo{pages}{214303} (\bibinfo{year}{2019}).

\bibitem[{\citenamefont{Korobov et~al.}(2017)\citenamefont{Korobov, Hilico, and
  Karr}}]{KoHiKa17}
\bibinfo{author}{\bibfnamefont{V.~I.} \bibnamefont{Korobov}},
  \bibinfo{author}{\bibfnamefont{L.}~\bibnamefont{Hilico}}, \bibnamefont{and}
  \bibinfo{author}{\bibfnamefont{J.~P.} \bibnamefont{Karr}},
  \bibinfo{journal}{Phys. Rev. Lett.} \textbf{\bibinfo{volume}{118}},
  \bibinfo{pages}{233001} (\bibinfo{year}{2017}).

\bibitem[{\citenamefont{Alighanbari et~al.}(2018)\citenamefont{Alighanbari,
  Hansen, Korobov, and Schiller}}]{AlHaKoSc18}
\bibinfo{author}{\bibfnamefont{S.}~\bibnamefont{Alighanbari}},
  \bibinfo{author}{\bibfnamefont{M.~G.} \bibnamefont{Hansen}},
  \bibinfo{author}{\bibfnamefont{V.~I.} \bibnamefont{Korobov}},
  \bibnamefont{and} \bibinfo{author}{\bibfnamefont{S.}~\bibnamefont{Schiller}},
  \bibinfo{journal}{Nature Physics} \textbf{\bibinfo{volume}{14}},
  \bibinfo{pages}{555} (\bibinfo{year}{2018}).

\bibitem[{\citenamefont{Wang and Yan}(2018)}]{WaYa18}
\bibinfo{author}{\bibfnamefont{L.~M.} \bibnamefont{Wang}} \bibnamefont{and}
  \bibinfo{author}{\bibfnamefont{Z.-C.} \bibnamefont{Yan}},
  \bibinfo{journal}{Phys. Rev. A} \textbf{\bibinfo{volume}{97}},
  \bibinfo{pages}{060501(R)} (\bibinfo{year}{2018}).

\bibitem[{\citenamefont{Puchalski et~al.}(2019)\citenamefont{Puchalski, Komasa,
  Czachorowski, and Pachucki}}]{PuKoCzPa19}
\bibinfo{author}{\bibfnamefont{M.}~\bibnamefont{Puchalski}},
  \bibinfo{author}{\bibfnamefont{J.}~\bibnamefont{Komasa}},
  \bibinfo{author}{\bibfnamefont{P.}~\bibnamefont{Czachorowski}},
  \bibnamefont{and} \bibinfo{author}{\bibfnamefont{K.}~\bibnamefont{Pachucki}},
  \bibinfo{journal}{Phys. Rev. Lett.} \textbf{\bibinfo{volume}{122}},
  \bibinfo{pages}{103003} (\bibinfo{year}{2019}).

\bibitem[{\citenamefont{H\"olsch et~al.}(2019)\citenamefont{H\"olsch, Beyer,
  Salumbides, Eikema, Ubachs, Jungen, and Merkt}}]{HoBeSaEiUbJuMe19}
\bibinfo{author}{\bibfnamefont{N.}~\bibnamefont{H\"olsch}},
  \bibinfo{author}{\bibfnamefont{M.}~\bibnamefont{Beyer}},
  \bibinfo{author}{\bibfnamefont{E.~J.} \bibnamefont{Salumbides}},
  \bibinfo{author}{\bibfnamefont{K.~S.} \bibnamefont{Eikema}},
  \bibinfo{author}{\bibfnamefont{W.}~\bibnamefont{Ubachs}},
  \bibinfo{author}{\bibfnamefont{C.}~\bibnamefont{Jungen}}, \bibnamefont{and}
  \bibinfo{author}{\bibfnamefont{F.}~\bibnamefont{Merkt}},
  \bibinfo{journal}{Phys. Rev. Lett.} \textbf{\bibinfo{volume}{122}},
  \bibinfo{pages}{103002} (\bibinfo{year}{2019}).

\bibitem[{\citenamefont{Tung et~al.}(2012)\citenamefont{Tung, Pavanello, and
  Adamowicz}}]{TuPaAd12}
\bibinfo{author}{\bibfnamefont{W.-C.} \bibnamefont{Tung}},
  \bibinfo{author}{\bibfnamefont{M.}~\bibnamefont{Pavanello}},
  \bibnamefont{and}
  \bibinfo{author}{\bibfnamefont{L.}~\bibnamefont{Adamowicz}},
  \bibinfo{journal}{J. Chem. Phys.} \textbf{\bibinfo{volume}{136}},
  \bibinfo{pages}{104309} (\bibinfo{year}{2012}).

\bibitem[{\citenamefont{Semeria et~al.}(2016)\citenamefont{Semeria, Jansen, and
  Merkt}}]{SeJaMe16}
\bibinfo{author}{\bibfnamefont{L.}~\bibnamefont{Semeria}},
  \bibinfo{author}{\bibfnamefont{P.}~\bibnamefont{Jansen}}, \bibnamefont{and}
  \bibinfo{author}{\bibfnamefont{F.}~\bibnamefont{Merkt}}, \bibinfo{journal}{J.
  Chem. Phys.} \textbf{\bibinfo{volume}{145}}, \bibinfo{pages}{204301}
  (\bibinfo{year}{2016}).

\bibitem[{\citenamefont{M\'atyus}(2018)}]{Ma18he2p}
\bibinfo{author}{\bibfnamefont{E.}~\bibnamefont{M\'atyus}},
  \bibinfo{journal}{J. Chem. Phys.} \textbf{\bibinfo{volume}{149}},
  \bibinfo{pages}{194112} (\bibinfo{year}{2018}).

\bibitem[{\citenamefont{Jansen et~al.}(2018)\citenamefont{Jansen, Semeria, and
  Merkt}}]{JaSeMe18}
\bibinfo{author}{\bibfnamefont{P.}~\bibnamefont{Jansen}},
  \bibinfo{author}{\bibfnamefont{L.}~\bibnamefont{Semeria}}, \bibnamefont{and}
  \bibinfo{author}{\bibfnamefont{F.}~\bibnamefont{Merkt}}, \bibinfo{journal}{J.
  Chem. Phys.} \textbf{\bibinfo{volume}{149}}, \bibinfo{pages}{154302}
  (\bibinfo{year}{2018}).

\bibitem[{\citenamefont{Stanke et~al.}(2009)\citenamefont{Stanke, Bubin, and
  Adamowicz}}]{StBuAd09}
\bibinfo{author}{\bibfnamefont{M.}~\bibnamefont{Stanke}},
  \bibinfo{author}{\bibfnamefont{S.}~\bibnamefont{Bubin}}, \bibnamefont{and}
  \bibinfo{author}{\bibfnamefont{L.}~\bibnamefont{Adamowicz}},
  \bibinfo{journal}{Phys. Rev. A} \textbf{\bibinfo{volume}{79}},
  \bibinfo{pages}{060501R} (\bibinfo{year}{2009}).

\bibitem[{\citenamefont{Herman and Asgharian}(1966)}]{HeAs66}
\bibinfo{author}{\bibfnamefont{R.~M.} \bibnamefont{Herman}} \bibnamefont{and}
  \bibinfo{author}{\bibfnamefont{A.}~\bibnamefont{Asgharian}},
  \bibinfo{journal}{J. Mol. Spectrosc.} \textbf{\bibinfo{volume}{19}},
  \bibinfo{pages}{305} (\bibinfo{year}{1966}).

\bibitem[{\citenamefont{Herman and Ogilvie}(1998)}]{HeOg98}
\bibinfo{author}{\bibfnamefont{R.~M.} \bibnamefont{Herman}} \bibnamefont{and}
  \bibinfo{author}{\bibfnamefont{J.~F.} \bibnamefont{Ogilvie}},
  \bibinfo{journal}{Adv. Chem. Phys.} \textbf{\bibinfo{volume}{103}},
  \bibinfo{pages}{187} (\bibinfo{year}{1998}).

\bibitem[{\citenamefont{Bunker and Moss}(1977)}]{BuMo77}
\bibinfo{author}{\bibfnamefont{P.~R.} \bibnamefont{Bunker}} \bibnamefont{and}
  \bibinfo{author}{\bibfnamefont{R.~E.} \bibnamefont{Moss}},
  \bibinfo{journal}{Mol. Phys.} \textbf{\bibinfo{volume}{33}},
  \bibinfo{pages}{417} (\bibinfo{year}{1977}).

\bibitem[{\citenamefont{Bunker and Moss}(1980)}]{BuMo80}
\bibinfo{author}{\bibfnamefont{P.~R.} \bibnamefont{Bunker}} \bibnamefont{and}
  \bibinfo{author}{\bibfnamefont{R.~E.} \bibnamefont{Moss}},
  \bibinfo{journal}{J. Mol. Spectrosc.} \textbf{\bibinfo{volume}{80}},
  \bibinfo{pages}{217} (\bibinfo{year}{1980}).

\bibitem[{\citenamefont{Pachucki and Komasa}(2009)}]{PaKo09}
\bibinfo{author}{\bibfnamefont{K.}~\bibnamefont{Pachucki}} \bibnamefont{and}
  \bibinfo{author}{\bibfnamefont{J.}~\bibnamefont{Komasa}},
  \bibinfo{journal}{J. Chem. Phys.} \textbf{\bibinfo{volume}{130}},
  \bibinfo{pages}{164113} (\bibinfo{year}{2009}).

\bibitem[{\citenamefont{Scherrer et~al.}(2017)\citenamefont{Scherrer, Agostini,
  Sebastiani, Gross, and Vuilleumier}}]{prx17}
\bibinfo{author}{\bibfnamefont{A.}~\bibnamefont{Scherrer}},
  \bibinfo{author}{\bibfnamefont{F.}~\bibnamefont{Agostini}},
  \bibinfo{author}{\bibfnamefont{D.}~\bibnamefont{Sebastiani}},
  \bibinfo{author}{\bibfnamefont{E.~K.~U.} \bibnamefont{Gross}},
  \bibnamefont{and}
  \bibinfo{author}{\bibfnamefont{R.}~\bibnamefont{Vuilleumier}},
  \bibinfo{journal}{Phys. Rev. X} \textbf{\bibinfo{volume}{7}},
  \bibinfo{pages}{031035} (\bibinfo{year}{2017}).

\bibitem[{\citenamefont{Schwenke}(2001)}]{Sch01H2O}
\bibinfo{author}{\bibfnamefont{D.~W.} \bibnamefont{Schwenke}},
  \bibinfo{journal}{J. Phys. Chem. A} \textbf{\bibinfo{volume}{105}},
  \bibinfo{pages}{2352} (\bibinfo{year}{2001}).

\bibitem[{\citenamefont{Przybytek et~al.}(2017)\citenamefont{Przybytek, Cencek,
  Jeziorski, and Szalewicz}}]{PrCeJeSz17}
\bibinfo{author}{\bibfnamefont{M.}~\bibnamefont{Przybytek}},
  \bibinfo{author}{\bibfnamefont{W.}~\bibnamefont{Cencek}},
  \bibinfo{author}{\bibfnamefont{B.}~\bibnamefont{Jeziorski}},
  \bibnamefont{and}
  \bibinfo{author}{\bibfnamefont{K.}~\bibnamefont{Szalewicz}},
  \bibinfo{journal}{Phys. Rev. Lett.} \textbf{\bibinfo{volume}{119}},
  \bibinfo{pages}{123401} (\bibinfo{year}{2017}).

\bibitem[{\citenamefont{M\'atyus and Teufel}(2019)}]{MaTe19}
\bibinfo{author}{\bibfnamefont{E.}~\bibnamefont{M\'atyus}} \bibnamefont{and}
  \bibinfo{author}{\bibfnamefont{S.}~\bibnamefont{Teufel}},
  \bibinfo{journal}{J. Chem. Phys.} \textbf{\bibinfo{volume}{150}},
  \bibinfo{pages}{214303} (\bibinfo{year}{2019}).

\bibitem[{\citenamefont{Muolo et~al.}(2018{\natexlab{a}})\citenamefont{Muolo,
  M\'atyus, and Reiher}}]{MuMaRe18proj}
\bibinfo{author}{\bibfnamefont{A.}~\bibnamefont{Muolo}},
  \bibinfo{author}{\bibfnamefont{E.}~\bibnamefont{M\'atyus}}, \bibnamefont{and}
  \bibinfo{author}{\bibfnamefont{M.}~\bibnamefont{Reiher}},
  \bibinfo{journal}{J. Chem. Phys.} \textbf{\bibinfo{volume}{149}},
  \bibinfo{pages}{184105} (\bibinfo{year}{2018}{\natexlab{a}}).

\bibitem[{\citenamefont{Bubin and Adamowicz}(2006)}]{BuAd06}
\bibinfo{author}{\bibfnamefont{S.}~\bibnamefont{Bubin}} \bibnamefont{and}
  \bibinfo{author}{\bibfnamefont{L.}~\bibnamefont{Adamowicz}},
  \bibinfo{journal}{J. Chem. Phys.} \textbf{\bibinfo{volume}{124}},
  \bibinfo{pages}{224317} (\bibinfo{year}{2006}).

\bibitem[{\citenamefont{Bubin and Adamowicz}(2008)}]{BuAd08}
\bibinfo{author}{\bibfnamefont{S.}~\bibnamefont{Bubin}} \bibnamefont{and}
  \bibinfo{author}{\bibfnamefont{L.}~\bibnamefont{Adamowicz}},
  \bibinfo{journal}{J. Chem. Phys.} \textbf{\bibinfo{volume}{128}},
  \bibinfo{pages}{114107} (\bibinfo{year}{2008}).

\bibitem[{\citenamefont{Suzuki and Varga}(1998)}]{SuVaBook98}
\bibinfo{author}{\bibfnamefont{Y.}~\bibnamefont{Suzuki}} \bibnamefont{and}
  \bibinfo{author}{\bibfnamefont{K.}~\bibnamefont{Varga}},
  \emph{\bibinfo{title}{Stochastic Variational Approach to Quantum-Mechanical
  Few-Body Problems}} (\bibinfo{publisher}{Springer-Verlag},
  \bibinfo{address}{Berlin}, \bibinfo{year}{1998}).

\bibitem[{\citenamefont{Mitroy et~al.}(2013)\citenamefont{Mitroy, Bubin,
  Horiuchi, Suzuki, Adamowicz, Cencek, Szalewicz, Komasa, Blume, and
  Varga}}]{rmp13}
\bibinfo{author}{\bibfnamefont{J.}~\bibnamefont{Mitroy}},
  \bibinfo{author}{\bibfnamefont{S.}~\bibnamefont{Bubin}},
  \bibinfo{author}{\bibfnamefont{W.}~\bibnamefont{Horiuchi}},
  \bibinfo{author}{\bibfnamefont{Y.}~\bibnamefont{Suzuki}},
  \bibinfo{author}{\bibfnamefont{L.}~\bibnamefont{Adamowicz}},
  \bibinfo{author}{\bibfnamefont{W.}~\bibnamefont{Cencek}},
  \bibinfo{author}{\bibfnamefont{K.}~\bibnamefont{Szalewicz}},
  \bibinfo{author}{\bibfnamefont{J.}~\bibnamefont{Komasa}},
  \bibinfo{author}{\bibfnamefont{D.}~\bibnamefont{Blume}}, \bibnamefont{and}
  \bibinfo{author}{\bibfnamefont{K.}~\bibnamefont{Varga}},
  \bibinfo{journal}{Rev. Mod. Phys.} \textbf{\bibinfo{volume}{85}},
  \bibinfo{pages}{693} (\bibinfo{year}{2013}).

\bibitem[{\citenamefont{Bubin et~al.}(2013)\citenamefont{Bubin, Pavanello,
  Tung, Sharkey, and Adamowicz}}]{chemrev13}
\bibinfo{author}{\bibfnamefont{S.}~\bibnamefont{Bubin}},
  \bibinfo{author}{\bibfnamefont{M.}~\bibnamefont{Pavanello}},
  \bibinfo{author}{\bibfnamefont{W.-C.} \bibnamefont{Tung}},
  \bibinfo{author}{\bibfnamefont{K.~L.} \bibnamefont{Sharkey}},
  \bibnamefont{and}
  \bibinfo{author}{\bibfnamefont{L.}~\bibnamefont{Adamowicz}},
  \bibinfo{journal}{Chem. Rev.} \textbf{\bibinfo{volume}{113}},
  \bibinfo{pages}{36} (\bibinfo{year}{2013}).

\bibitem[{spe()}]{specacc}
\bibinfo{note}{Spectroscopic accuracy is generally defined as obtaining
  (ro)vibrational state energies within better than a 1~\cm\ uncertainty.}

\bibitem[{\citenamefont{M\'atyus and Reiher}(2012)}]{MaRe12}
\bibinfo{author}{\bibfnamefont{E.}~\bibnamefont{M\'atyus}} \bibnamefont{and}
  \bibinfo{author}{\bibfnamefont{M.}~\bibnamefont{Reiher}},
  \bibinfo{journal}{J. Chem. Phys.} \textbf{\bibinfo{volume}{137}},
  \bibinfo{pages}{024104} (\bibinfo{year}{2012}).

\bibitem[{\citenamefont{Lindsay and McCall}(2001)}]{LiMc01}
\bibinfo{author}{\bibfnamefont{C.~M.} \bibnamefont{Lindsay}} \bibnamefont{and}
  \bibinfo{author}{\bibfnamefont{B.~J.} \bibnamefont{McCall}},
  \bibinfo{journal}{J. Mol. Spectrosc.} \textbf{\bibinfo{volume}{210}},
  \bibinfo{pages}{60} (\bibinfo{year}{2001}).

\bibitem[{\citenamefont{Bunker and Jensen}(1998)}]{BuJe98}
\bibinfo{author}{\bibfnamefont{P.~R.} \bibnamefont{Bunker}} \bibnamefont{and}
  \bibinfo{author}{\bibfnamefont{P.}~\bibnamefont{Jensen}},
  \emph{\bibinfo{title}{Molecular symmetry and spectroscopy, 2nd Edition}}
  (\bibinfo{publisher}{NRC Research Press}, \bibinfo{address}{Ottawa},
  \bibinfo{year}{1998}).

\bibitem[{\citenamefont{M\'atyus
  et~al.}(2011{\natexlab{a}})\citenamefont{M\'atyus, Hutter, M\"uller-Herold,
  and Reiher}}]{MaHuMuRe11a}
\bibinfo{author}{\bibfnamefont{E.}~\bibnamefont{M\'atyus}},
  \bibinfo{author}{\bibfnamefont{J.}~\bibnamefont{Hutter}},
  \bibinfo{author}{\bibfnamefont{U.}~\bibnamefont{M\"uller-Herold}},
  \bibnamefont{and} \bibinfo{author}{\bibfnamefont{M.}~\bibnamefont{Reiher}},
  \bibinfo{journal}{Phys. Rev. A} \textbf{\bibinfo{volume}{83}},
  \bibinfo{pages}{052512} (\bibinfo{year}{2011}{\natexlab{a}}).

\bibitem[{\citenamefont{M\'atyus
  et~al.}(2011{\natexlab{b}})\citenamefont{M\'atyus, Hutter, M\"uller-Herold,
  and Reiher}}]{MaHuMuRe11b}
\bibinfo{author}{\bibfnamefont{E.}~\bibnamefont{M\'atyus}},
  \bibinfo{author}{\bibfnamefont{J.}~\bibnamefont{Hutter}},
  \bibinfo{author}{\bibfnamefont{U.}~\bibnamefont{M\"uller-Herold}},
  \bibnamefont{and} \bibinfo{author}{\bibfnamefont{M.}~\bibnamefont{Reiher}},
  \bibinfo{journal}{J. Chem. Phys.} \textbf{\bibinfo{volume}{135}},
  \bibinfo{pages}{204302} (\bibinfo{year}{2011}{\natexlab{b}}).

\bibitem[{\citenamefont{Schild}(2019)}]{Sc19}
\bibinfo{author}{\bibfnamefont{A.}~\bibnamefont{Schild}},
  \bibinfo{journal}{Front. Chem. doi:10.3389/fchem.2019.00424}
  (\bibinfo{year}{2019}).

\bibitem[{\citenamefont{Suzuki et~al.}(1998)\citenamefont{Suzuki, Usukura, and
  Varga}}]{SuUsVa98}
\bibinfo{author}{\bibfnamefont{Y.}~\bibnamefont{Suzuki}},
  \bibinfo{author}{\bibfnamefont{J.}~\bibnamefont{Usukura}}, \bibnamefont{and}
  \bibinfo{author}{\bibfnamefont{K.}~\bibnamefont{Varga}}, \bibinfo{journal}{J.
  Phys. B: At. Mol. Opt. Phys.} \textbf{\bibinfo{volume}{31}},
  \bibinfo{pages}{31} (\bibinfo{year}{1998}).

\bibitem[{\citenamefont{Armour et~al.}(2005)\citenamefont{Armour, Richard, and
  Varga}}]{ArRiVa05}
\bibinfo{author}{\bibfnamefont{E.~A.~G.} \bibnamefont{Armour}},
  \bibinfo{author}{\bibfnamefont{J.-M.} \bibnamefont{Richard}},
  \bibnamefont{and} \bibinfo{author}{\bibfnamefont{K.}~\bibnamefont{Varga}},
  \bibinfo{journal}{Phys. Rep.} \textbf{\bibinfo{volume}{413}},
  \bibinfo{pages}{1} (\bibinfo{year}{2005}).

\bibitem[{\citenamefont{Varga et~al.}(1998{\natexlab{a}})\citenamefont{Varga,
  Usukura, and Suzuki}}]{VaUsSu98}
\bibinfo{author}{\bibfnamefont{K.}~\bibnamefont{Varga}},
  \bibinfo{author}{\bibfnamefont{J.}~\bibnamefont{Usukura}}, \bibnamefont{and}
  \bibinfo{author}{\bibfnamefont{Y.}~\bibnamefont{Suzuki}},
  \bibinfo{journal}{Phys. Rev. Lett.} \textbf{\bibinfo{volume}{80}},
  \bibinfo{pages}{1876} (\bibinfo{year}{1998}{\natexlab{a}}).

\bibitem[{\citenamefont{Mátyus}(2013)}]{Ma13}
\bibinfo{author}{\bibfnamefont{E.}~\bibnamefont{Mátyus}}, \bibinfo{journal}{J.
  Phys. Chem. A} \textbf{\bibinfo{volume}{117}}, \bibinfo{pages}{7195}
  (\bibinfo{year}{2013}).

\bibitem[{FeM()}]{FeMa19a}
\bibinfo{note}{D. Ferenc and E. M\'atyus, Precise computation of rovibronic
  resonances of molecular hydrogen: $EF~^1\Sigma_\text{g}^+$ inner-well
  rotational states. arXiv:1904.08609}.

\bibitem[{pes()}]{pesvalues}
\bibinfo{note}{These reference values have been computed in our earlier work
  \cite{MuMaRe18proj} using the GENIUSH program
  \cite{MaCzCs09,FaMaCs11,MaSzCs14} with the Polyansky--Tennyson \cite{PoTe99}
  mass model and the GLH3P PES \cite{PaAdAl12jcp} by switching off the
  relativistic corrections.}

\bibitem[{\citenamefont{Kinghorn and Adamowicz}(1999)}]{KiAd99}
\bibinfo{author}{\bibfnamefont{D.~B.} \bibnamefont{Kinghorn}} \bibnamefont{and}
  \bibinfo{author}{\bibfnamefont{L.}~\bibnamefont{Adamowicz}},
  \bibinfo{journal}{J. Chem. Phys.} \textbf{\bibinfo{volume}{110}},
  \bibinfo{pages}{7166} (\bibinfo{year}{1999}).

\bibitem[{\citenamefont{Bubin and Adamowicz}(2003)}]{BuAd03}
\bibinfo{author}{\bibfnamefont{S.}~\bibnamefont{Bubin}} \bibnamefont{and}
  \bibinfo{author}{\bibfnamefont{L.}~\bibnamefont{Adamowicz}},
  \bibinfo{journal}{J. Chem. Phys.} \textbf{\bibinfo{volume}{118}},
  \bibinfo{pages}{3079} (\bibinfo{year}{2003}).

\bibitem[{\citenamefont{Cafiero et~al.}(2003)\citenamefont{Cafiero, Bubin, and
  Adamowicz}}]{CaBuAd03}
\bibinfo{author}{\bibfnamefont{M.}~\bibnamefont{Cafiero}},
  \bibinfo{author}{\bibfnamefont{S.}~\bibnamefont{Bubin}}, \bibnamefont{and}
  \bibinfo{author}{\bibfnamefont{L.}~\bibnamefont{Adamowicz}},
  \bibinfo{journal}{Phys. Chem. Chem. Phys.} \textbf{\bibinfo{volume}{5}},
  \bibinfo{pages}{1491} (\bibinfo{year}{2003}).

\bibitem[{\citenamefont{Bednarz et~al.}(2005)\citenamefont{Bednarz, Bubin, and
  Adamowicz}}]{BeBuAd05}
\bibinfo{author}{\bibfnamefont{E.}~\bibnamefont{Bednarz}},
  \bibinfo{author}{\bibfnamefont{S.}~\bibnamefont{Bubin}}, \bibnamefont{and}
  \bibinfo{author}{\bibfnamefont{L.}~\bibnamefont{Adamowicz}},
  \bibinfo{journal}{Mol. Phys.} \textbf{\bibinfo{volume}{103}},
  \bibinfo{pages}{1169} (\bibinfo{year}{2005}).

\bibitem[{\citenamefont{Bubin et~al.}(2016)\citenamefont{Bubin, Formanek, and
  Adamowicz}}]{BuFoAd16}
\bibinfo{author}{\bibfnamefont{S.}~\bibnamefont{Bubin}},
  \bibinfo{author}{\bibfnamefont{M.}~\bibnamefont{Formanek}}, \bibnamefont{and}
  \bibinfo{author}{\bibfnamefont{L.}~\bibnamefont{Adamowicz}},
  \bibinfo{journal}{Chem. Phys. Lett.} \textbf{\bibinfo{volume}{647}},
  \bibinfo{pages}{122} (\bibinfo{year}{2016}).

\bibitem[{\citenamefont{Varga}(2019)}]{Va19}
\bibinfo{author}{\bibfnamefont{K.}~\bibnamefont{Varga}},
  \bibinfo{journal}{Phys. Rev. A} \textbf{\bibinfo{volume}{99}},
  \bibinfo{pages}{012504} (\bibinfo{year}{2019}).

\bibitem[{\citenamefont{Simmen et~al.}(2013)\citenamefont{Simmen, M\'atyus, and
  Reiher}}]{SiMaRe13}
\bibinfo{author}{\bibfnamefont{B.}~\bibnamefont{Simmen}},
  \bibinfo{author}{\bibfnamefont{E.}~\bibnamefont{M\'atyus}}, \bibnamefont{and}
  \bibinfo{author}{\bibfnamefont{M.}~\bibnamefont{Reiher}},
  \bibinfo{journal}{Mol. Phys.} \textbf{\bibinfo{volume}{111}},
  \bibinfo{pages}{2086} (\bibinfo{year}{2013}).

\bibitem[{\citenamefont{Muolo et~al.}(2018{\natexlab{b}})\citenamefont{Muolo,
  M\'atyus, and Reiher}}]{MuMaRe18}
\bibinfo{author}{\bibfnamefont{A.}~\bibnamefont{Muolo}},
  \bibinfo{author}{\bibfnamefont{E.}~\bibnamefont{M\'atyus}}, \bibnamefont{and}
  \bibinfo{author}{\bibfnamefont{M.}~\bibnamefont{Reiher}},
  \bibinfo{journal}{J. Chem. Phys.} \textbf{\bibinfo{volume}{148}},
  \bibinfo{pages}{084112} (\bibinfo{year}{2018}{\natexlab{b}}).

\bibitem[{\citenamefont{Varga et~al.}(1998{\natexlab{b}})\citenamefont{Varga,
  Suzuki, and Usukura}}]{VaSuUs98}
\bibinfo{author}{\bibfnamefont{K.}~\bibnamefont{Varga}},
  \bibinfo{author}{\bibfnamefont{Y.}~\bibnamefont{Suzuki}}, \bibnamefont{and}
  \bibinfo{author}{\bibfnamefont{J.}~\bibnamefont{Usukura}},
  \bibinfo{journal}{Few-Body Systems} \textbf{\bibinfo{volume}{24}},
  \bibinfo{pages}{81} (\bibinfo{year}{1998}{\natexlab{b}}).

\bibitem[{\citenamefont{Polizzi}(2009)}]{Po09}
\bibinfo{author}{\bibfnamefont{E.}~\bibnamefont{Polizzi}},
  \bibinfo{journal}{Phys. Rev. B} \textbf{\bibinfo{volume}{79}},
  \bibinfo{pages}{115112} (\bibinfo{year}{2009}).

\bibitem[{\citenamefont{M\'atyus et~al.}(2009)\citenamefont{M\'atyus, Czak\'o,
  and Cs\'asz\'ar}}]{MaCzCs09}
\bibinfo{author}{\bibfnamefont{E.}~\bibnamefont{M\'atyus}},
  \bibinfo{author}{\bibfnamefont{G.}~\bibnamefont{Czak\'o}}, \bibnamefont{and}
  \bibinfo{author}{\bibfnamefont{A.~G.} \bibnamefont{Cs\'asz\'ar}},
  \bibinfo{journal}{J. Chem. Phys.} \textbf{\bibinfo{volume}{130}},
  \bibinfo{pages}{134112} (\bibinfo{year}{2009}).

\bibitem[{\citenamefont{F\'abri et~al.}(2011)\citenamefont{F\'abri, M\'atyus,
  and Cs\'asz\'ar}}]{FaMaCs11}
\bibinfo{author}{\bibfnamefont{C.}~\bibnamefont{F\'abri}},
  \bibinfo{author}{\bibfnamefont{E.}~\bibnamefont{M\'atyus}}, \bibnamefont{and}
  \bibinfo{author}{\bibfnamefont{A.~G.} \bibnamefont{Cs\'asz\'ar}},
  \bibinfo{journal}{J. Chem. Phys.} \textbf{\bibinfo{volume}{134}},
  \bibinfo{pages}{074105} (\bibinfo{year}{2011}).

\bibitem[{\citenamefont{M{\'a}tyus et~al.}(2014)\citenamefont{M{\'a}tyus,
  Szidarovszky, and Cs{\'a}sz{\'a}r}}]{MaSzCs14}
\bibinfo{author}{\bibfnamefont{E.}~\bibnamefont{M{\'a}tyus}},
  \bibinfo{author}{\bibfnamefont{T.}~\bibnamefont{Szidarovszky}},
  \bibnamefont{and} \bibinfo{author}{\bibfnamefont{A.~G.}
  \bibnamefont{Cs{\'a}sz{\'a}r}}, \bibinfo{journal}{J. Chem. Phys.}
  \textbf{\bibinfo{volume}{141}}, \bibinfo{pages}{154111}
  (\bibinfo{year}{2014}).

\bibitem[{\citenamefont{Polyansky and Tennyson}(1999)}]{PoTe99}
\bibinfo{author}{\bibfnamefont{O.~L.} \bibnamefont{Polyansky}}
  \bibnamefont{and} \bibinfo{author}{\bibfnamefont{J.}~\bibnamefont{Tennyson}},
  \bibinfo{journal}{J. Chem. Phys.} \textbf{\bibinfo{volume}{110}},
  \bibinfo{pages}{5056} (\bibinfo{year}{1999}).

\bibitem[{\citenamefont{Pavanello et~al.}(2012)\citenamefont{Pavanello,
  Adamowicz, Alijah, Zobov, Mizus, Polyansky, Tennyson, Szidarovszky, and
  Cs\'asz\'ar}}]{PaAdAl12jcp}
\bibinfo{author}{\bibfnamefont{M.}~\bibnamefont{Pavanello}},
  \bibinfo{author}{\bibfnamefont{L.}~\bibnamefont{Adamowicz}},
  \bibinfo{author}{\bibfnamefont{A.}~\bibnamefont{Alijah}},
  \bibinfo{author}{\bibfnamefont{N.~F.} \bibnamefont{Zobov}},
  \bibinfo{author}{\bibfnamefont{I.}~\bibnamefont{Mizus}},
  \bibinfo{author}{\bibfnamefont{O.~L.} \bibnamefont{Polyansky}},
  \bibinfo{author}{\bibfnamefont{J.}~\bibnamefont{Tennyson}},
  \bibinfo{author}{\bibfnamefont{T.}~\bibnamefont{Szidarovszky}},
  \bibnamefont{and} \bibinfo{author}{\bibfnamefont{A.~G.}
  \bibnamefont{Cs\'asz\'ar}}, \bibinfo{journal}{J. Chem. Phys.}
  \textbf{\bibinfo{volume}{136}}, \bibinfo{pages}{184303}
  (\bibinfo{year}{2012}).

\end{thebibliography}

\end{document}